\documentclass[aps,prl,showpacs,twocolumn,superscriptaddress]{revtex4}
\usepackage{bm,color}
\usepackage{amsmath, amssymb}
\usepackage{graphicx}
\usepackage{braket}
\usepackage{color}
\usepackage{comment}

\begin{document}
%\draft
\title {Predicted novel type of photoinduced topological phase transition accompanied by collision and collapse of Dirac-cone pair in organic salt $\alpha$-(BEDT-TTF)$_2$I$_3$}
\author{Keisuke Kitayama}
\affiliation{Department of Physics, University of Tokyo, Hongo, Bunkyo-ku, Tokyo 113-8656, Japan}
\author{Masao Ogata}
\affiliation{Department of Physics, University of Tokyo, Hongo, Bunkyo-ku, Tokyo 113-8656, Japan}
\affiliation{Trans-scale Quantum Science Institute, University of Tokyo, Bunkyo-ku, Tokyo 113-0033, Japan}
\author{Masahito Mochizuki}
\affiliation{Department of Applied Physics, Waseda University, Okubo, Shinjuku-ku, Tokyo 169-8555, Japan}
\author{Yasuhiro Tanaka}
\affiliation{Department of Applied Physics, Waseda University, Okubo, Shinjuku-ku, Tokyo 169-8555, Japan}
\begin{abstract}
Photoinduced topological phase transitions in the Dirac-electron systems have attracted intensive research interest since its theoretical prediction in graphene, where the application of circularly polarized light opens a gap at the Dirac points and renders the system a topologically nontrivial Chern insulator phase through breaking the time-reversal symmetry. However, most of the previously studied phenomena in two-dimensional Dirac systems are basically based on the same physicical mechanism, i.e., the gap opening in the Dirac electron bands with circularly polarized light, and it is lacking in variety of the physics. In this paper, we theoretically predict a novel type of photoinduced topological phase transition accompanied by collision and collapse of gapped Dirac points in the organic salt $\alpha$-(BEDT-TTF)$_2$I$_3$. By constructing the Floquet theory for this compound, we demonstrate that the irradiation of elliptically polarized light causes collision of the Dirac points through the photoinduced band deformation and their collapse, which eventually results in the topological phase transition from a topological semimetal with gapped Dirac cones to a normal insulator when the elliptical axis is oriented at a specfic angle with respect to the crystallographic axes. We argue that this novel photoinduced phase transition can be experimentally detected by the measurement of Hall conductivity. The present work enriches the fundamental physics of photoinduced topological phase transitions and thus contribute to development of this rapidly growing research field.
\end{abstract}
\maketitle
%\sloppy \maketitle

\section{Introduction}
%%%%%%%%%%%%%%%%%%%%%%%%%%%%%%%%
\begin{figure}[tb]
\includegraphics[scale=1.0]{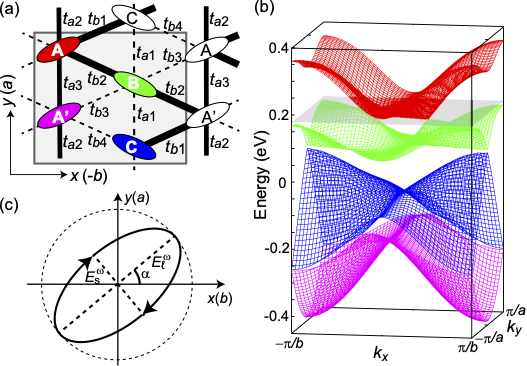}
\caption{(a) BEDT-TTF layer in $\alpha$-(BEDT-TTF)$_2$I$_3$. The unit cell (the gray rectangle) contains four molecular sublattices (A, A$^\prime$, B, C) and seven kinds of transfer integrals. (b) Band structure with a pair of tilted Dirac cones between the valence band (the third band) and the conduction band (the fourth band) in $\alpha$-(BEDT-TTF)$_2$I$_3$ obtained by diagonalizing the tight-binding Hamiltonian in Eq.~(\ref{eq:TBm1}). (c) Elliptically polarized light applied to the  BEDT-TTF layer, which is characterized by the short-axis and long-axis amplitudes ($E^\omega_{\rm s}$, $E^\omega_{\rm \ell}$), the polarization angle $\alpha$, and the frequency $\hbar\omega$.}
\label{Fig01}
\end{figure}
%%%%%%%%%%%%%%%%%%%%%%%%%%%%%%%%
Research on the photoinduced topological phase transitions has been launched by a prescient theoretical work in Refs.~\cite{Oka09,Kitagawa11} that predicted the emergence of Chern insulator phase~\cite{Haldane88} in graphene irradiated by circularly polarized light. The research has been accelerated recently by rapid development of the laser technology~\cite{Yonemitsu06,Tokura06,Bukov15,Aoki14,Basov17}. Indeed, laser-induced Hall currents as an implication of emergent topological phase have been observed experimentally in photodriven graphene~\cite{McIver19}, and possible relevance to the predicted photoinduced Chern insulator phase has been discussed intensively~\cite{Sato19a,Sato19b,Nuske20}.

Since the pioneering work in Refs.~\cite{Oka09,Kitagawa11}, many theoretical studies of photoinduced topological phase transitions have been conducted by using the Floquet theory~\cite{Kitagawa10,Inoue10,Lindner11,Rudner13,Grushin14,ZhengW14,JYZou16,Ezawa13,Kang20,Claassen16,Hubener17,ZhangMY19,ZYan16,Sato16,Kitamura17,LDu17,Ezawa17,Takasan15,Takasan17a,Takasan17b,Menon18,ChenR18,Tanaka20,Kitayama20,Kitayama21a,Tanaka21}. The photoinduced phase transitions argued in these studies can be classified into two categories, i.e., those induced by off-resonant light and those induced by on-resonant light~\cite{Rudner20,Oka19,Giovannini19}. The resonant drives can cause a band inversion in gapped systems such as band insulators and semiconductors~\cite{Lindner11}. On the other hand, the off-resonant drives can change the topological properties in gapless Dirac and Weyl electron systems like graphene~\cite{Oka09,Kitagawa11}. In the latter case, an effective static Hamiltonian (called Floquet Hamiltonian) derived using the Floquet theorem is commonly employed~\cite{Kitagawa11,Mikami16}. 

Indeed, the theoretical studies based on the Floquet Hamiltonians have predicted several nontrivial photoinduced phases emerging from the Dirac semimetal phases under irradiation with high-frequency circularly polarized light. For example, the Floquet Weyl semimetals are predicted to come up from the three-dimensional Dirac semimetals, and the Floquet Chern insulators are from the two-dimensional Dirac semimetals~\cite{Oka09,Inoue10,Kitagawa11,Hubener17}. In particular, the photoinduced Chern insulator phases in two dimensions have been intensively studied so far. However, their phase transitions are basically based on the same physical mechanism, i.e., the gap opening at the Dirac points through breaking the time-reversal symmetry with circularly polarized light~\cite{Mikami16, Oka09, Kitagawa11, Piskunow14, Usaj14}.

Recently, possible photoinduced phase transitions in the organic salt $\alpha$-(BEDT-TTF)$_2$I$_3$ have been theoretically studied using the Floquet theory~\cite{Kitayama20, Kitayama21a, Kitayama21b}. This compound is composed of stacked two-dimensional conducting layers, unit cell of which contains four BEDT-TTF molecules named A, A$^\prime$, B and C [Fig. \ref{Fig01}(a)]. At ambient pressure, this compound has a charge-ordered ground state, but it vanishes when we apply a uniaxial pressure of $P_a > 4$ kbar. In the absence of charge order, a pair of tilted Dirac-cone bands appear between the third and fourth bands among four bands originating from the four molecules in the unit cell [Fig. \ref{Fig01}(b)]~\cite{Tajima06, Kajita14, Katayama06, Kobayashi07}. Moreover, these Dirac cones are gapless, and their Dirac points are located at the Fermi level because the electron filling is $n_{\rm e}$=3/4. This kind of Dirac semimetal phase is expected to host fascinating topological properties and phenomena~\cite{Suzumura11,Osada17,Kitayama20,Kitayama21a,Kitayama21b,Osada21,Fujiyama22,Ogata22}. Recent theoretical studies revealed that irradiation with circularly polarized light gives rise to the Chern insulator phase with gapped Dirac-cone bands through breaking the time-reversal symmetry~\cite{Kitayama20}, while irradiation with linearly polarized light gives rise to the pair annihilation of Dirac points with opposite topological charges~\cite{Kitayama21a}.

In fact, the organic compound $\alpha$-(BEDT-TTF)$_2$I$_3$ is a special material that provides us a rare opportunity to explore a variety of photoinduced phase-transition phenomena for several reasons~\cite{Kitayama20, Kitayama21a, Kitayama21b}. First, effects of light electric field, which are incorporated by  Peierls phases on the transfer integrals, are enhanced because this molecular compound has large lattice constants and the Peierls phases are scaled with the vector potential multiplied by the lattice constants. Second, a set of four bands associated with the BEDT-TTF molecules are located around the Fermi level and are well-separated from upper and lower band sets, which guarantees a presence of light-frequency window that realizes the off-resonant condition. Lastly but most importantly, this compound has a relatively delicate band structure described by seven kinds of transfer integrals in real space. Because of this, various patterns of dynamical band deformations are possible depending on the light polarizations, which can give rise to a variety of photoinduced phenomena and phases. Indeed, the above-mentioned pair annihilation of Dirac points in Ref.~\cite{Kitayama21a} is expected for linearly polarized light with a specific polarization angle where the light electric field is nearly parallel to [110] direction. Because of these perculiarities, it is an issue of interest to explore novel photoinduced phenomena in this compound by varying several parameters of light, e.g., type and direction of polarization, intensity, frequency, and even duration.

\begin{figure}[tb]
\includegraphics[scale=1.0]{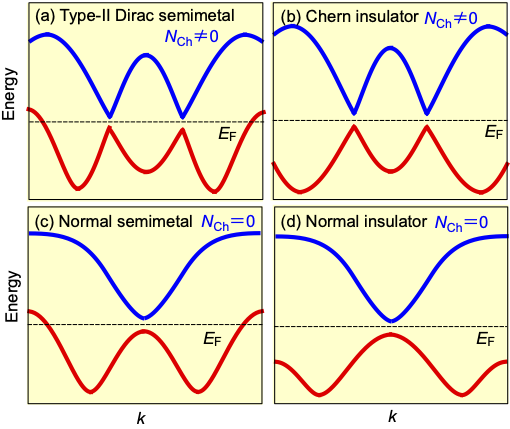}
\caption{Schematic band structures of various photoinduced phases discussed in this paper, i.e., (a) topological semimetal with gapped Dirac cones, (b) Chern insulator, (c) normal semimetal, and (d) normal insulator. The former two are topological phases, while the latter two are nontopological phases.}
\label{Fig02}
\end{figure}

In this paper, we theoretically predict a novel type of photoinduced topological phase transition accompanied by collision and collapse of two massive Dirac cones in the organic salt $\alpha$-(BEDT-TTF)$_2$I$_3$. By constructing the Floquet theory of this compound, we demonstrate that irradiation with elliptically polarized light with a specific elliptical-axis angle [Fig.~\ref{Fig01}(c)] causes collision of the Dirac points and resulting their collapse through dynamical band deformation, which leads to the topological phase transition from a semimetal with gapped Dirac cones (topological) to a normal insulator (nontopological). This novel type of topological phase transition is distinct from the photoinduced topological phase transition shown in our previous paper that occurs in $\alpha$-(BEDT-TTF)$_2$I$_3$ irradiated with elliptically polarized light~\cite{Kitayama21b}. We also elucidate a rich nonequilibrium phase diagram in plane of the amplitude and elliptical-axis angle of light that contains four phases: topological semimetal phase, Chern insulator phase, normal semimetal phase, and normal insulator phase [Fig.~\ref{Fig02}]. In addition, we propose that the Hall conductivity can be a promising probe of the predicted unique photoinduced phase transition. The present work will advance the research field of photoinduced topological phase transitions through enriching their fundamental physics.

\section{Model and Methods}
%%%%%%%%%%%%%%%%%%%%%%%%%%%%%%%%
%%%%%%%%%%%%%%%%%%%%%%%%%%%%%%%%
We start with a tight-binding model that describes the electronic structure in the BEDT-TTF layer of the organic conductor $\alpha$-(BEDT-TTF)$_2$I$_3$ before light irradiation~\cite{Katayama06,Kajita14}. The tight-binding model is given by,
%%%%%%%%%%%%%%%%%%%%%%%%%
\begin{eqnarray}
H = \sum_{\langle i,j \rangle}\sum_{\alpha, \beta} 
t_{i\alpha,j\beta}c_{i,\alpha}^{\dagger}c_{j,\beta},
\label{eq:TBm1}
\end{eqnarray}
%%%%%%%%%%%%%%%%%%%%%%%%%
where $i$ and $j$ are indices of the unit cells, whereas $\alpha$ and $\beta$ are indices of the molecular sites (A, A$^\prime$, B and C). The symbol $c_{j,\beta}^{\dagger}$ ($c_{i,\alpha}$) represents the electron creation (annihilation) operator, and $t_{i\alpha,j\beta}$ denotes transfer integrals between adjacent molecular sites. We use the following formulas of the pressure-dependent transfer integrals proposed in Ref.~\cite{Kobayashi04}, $t_{a1} = -0.028(1 + 0.089P_a)$ eV, $t_{a2} = 0.048(1 + 0.167P_a)$ eV, $t_{a3} = -0.020(1 - 0.025P_a)$ eV, $t_{b1} = 0.123$ eV, $t_{b2} = 0.140(1 + 0.011P_a)$ eV, $t_{b3} = -0.062(1 + 0.032P_a)$ eV, and $t_{b4} = -0.025$ eV, where $P_a$ is the value of uniaxial pressure in kbar units~\cite{Kobayashi04}. We set $P_a = 4$ kbar in the calculations.
%%%%%%%%%%%%%%%%%%%%%%%%%
%\begin{align}
%&t_{a1} = -0.028(1 + 0.089P_a) \;{\rm eV},\nonumber \\
%&t_{a2} = 0.048(1 + 0.167P_a) \;{\rm eV},\nonumber \\
%&t_{a3} = -0.020(1 - 0.025P_a) \;{\rm eV},\nonumber \\
%&t_{b1} = 0.123 \;{\rm eV},\nonumber \\
%&t_{b2} = 0.140(1 + 0.011P_a) \;{\rm eV},\nonumber \\
%&t_{b3} = -0.062(1 + 0.032P_a) \;{\rm eV},\nonumber \\
%&t_{b4} = -0.025 \;{\rm eV}.\nonumber
%\end{align}
%%%%%%%%%%%%%%%%%%%%%%%%%

We consider the situation that $\alpha$-(BEDT-TTF)$_2$I$_3$ is irradiated with elliptically polarized light, whose vector potential is given by,
%%%%%%%%%%%%%%%%%%%%%%%%%
\begin{eqnarray}
\bm{A}(\tau) &=&\left(
\begin{array}{cc}
\cos\alpha & -\sin\alpha \\
\sin\alpha & \cos\alpha
\end{array}
\right)	
\left(
\begin{array}{c}
A_{\rm \ell}\cos(\omega\tau) \\
A_{\rm s}\sin(\omega\tau)
\end{array}
\right),
\label{eq:vecpot}
\end{eqnarray}
%%%%%%%%%%%%%%%%%%%%%%%%%
where $\omega$ is the frequency of light. This vector potential produces a time-dependent electric field,
%%%%%%%%%%%%%%%%%%%%%%%%%
\begin{eqnarray}
\bm E(\tau)&=&-\frac{d\bm A(\tau)}{d\tau} \nonumber \\
&=&\left(
\begin{array}{cc}
\cos\alpha & -\sin\alpha \\
\sin\alpha & \cos\alpha
\end{array}
\right)	
\left(
\begin{array}{c}
-E_{\rm \ell}^\omega\sin(\omega\tau) \\
E_{\rm s}^\omega\cos(\omega\tau)
\end{array}
\right),
\label{eq:acEfld}
\end{eqnarray}
%%%%%%%%%%%%%%%%%%%%%%%%%
where the amplitude of light is defined as $E_\gamma^\omega = A_\gamma\omega$ ($\gamma$=s, $\ell$). Equation~(\ref{eq:acEfld}) describes elliptically polarized light in Fig.~\ref{Fig01}(c), which is characterized by the long-axis amplitude $E^\omega_{\rm \ell}$, the short-axis amplitude $E^\omega_{\rm s}$, the elliptical-axis angle $\alpha$, and the frequency $\hbar\omega$. Note that we introduce the elliptical-axis angle to investigate the collapse and collision of Dirac points in this paper, which is distinct from the elliptically polarized light considered in our previous paper~\cite{Kitayama21b}. The formalism in Ref.~\cite{Kitayama21b} corresponds to the case of $\alpha = 0$.

The effects of light irradiation are taken into account by attaching the Peierls phases to the transfer integrals in Eq.~(\ref{eq:TBm1}). Accordingly, the time-dependent tight-binding Hamiltonian for the photodriven $\alpha$-(BEDT-TTF)$_2$I$_3$ is given by,~\cite{Yonemitsu06,Kitayama20,Tanaka10,Miyashita10}
%%%%%%%%%%%%%%%%%%%%%%%%%
\begin{eqnarray}
H(\tau)&=&\sum_{\langle i,j \rangle}\sum_{\alpha, \beta} t_{i\alpha,j\beta} \;
e^{-ie\bm{A}(\tau)\cdot(\bm{r}_{i\alpha} - \bm{r}_{j\beta})/\hbar}
c_{i,\alpha}^{\dagger}c_{j,\beta}.
\label{eq:TBm2}
\end{eqnarray}
%%%%%%%%%%%%%%%%%%%%%%%%%
%%%%%%%%%%%%%%%%%%%%%%%%%
%\begin{eqnarray}
%H(\tau)=\sum_{\langle i,j \rangle}\sum_{\alpha, \beta} t_{i\alpha,j\beta} \;
%\exp\left\{-i\frac{e}{\hbar}\bm{A}(\tau)\cdot(\bm{r}_{i\alpha} - \bm{r}_{j\beta})\right\}
%c_{i,\alpha}^{\dagger}c_{j,\beta},
%\label{eq:TBm2}
%\end{eqnarray}
%%%%%%%%%%%%%%%%%%%%%%%%%
Here $\bm r_{i\alpha}=(b\tilde{x}_{i\alpha}, a\tilde{y}_{i\alpha})$ is the spatial coordinates of the $\alpha$th molecular sublattice in the $i$th unit cell, where $a$(=0.9187 nm) and $b$(=1.0793 nm) are the lattice constants along the $y$ and $x$ axes, respectively~\cite{Mori12}.

\begin{comment}
Time evolutions of the electrons in photodriven $\alpha$-(BEDT-TTF)$_2$I$_3$ are described by the time-dependent Schr\"{o}dinger equation,
%%%%%%%%%%%%%%%%%%%%%%%%%
\begin{eqnarray}
i \hbar \frac{\partial}{\partial \tau} \ket{\Psi(\tau)} = H(\tau)\ket{\Psi(\tau)}. 
\label{eq:Scheq}
\end{eqnarray}
%%%%%%%%%%%%%%%%%%%%%%%%%
The Hamiltonian $H(\tau)$ is time periodic as $H(\tau)=H(\tau+T)$, with a temporal periodicity of the light $T (= 2\pi / \omega)$. The wavefunction $\ket{\Psi(\tau)}$ in such systems can be represented in the form
%%%%%%%%%%%%%%%%%%%%%%%%%
\begin{eqnarray}
\ket{\Psi(\tau)} = e^{i\varepsilon \tau/\hbar}\ket{\Phi(\tau)},
\label{eq:Flqthm}
\end{eqnarray}
%%%%%%%%%%%%%%%%%%%%%%%%%
where $\varepsilon$ is referred to as quasienergy. The state $\ket{\Phi(\tau)}$ satisfies $\ket{\Phi(\tau)}=\ket{\Phi(\tau +T)}$. This theorem is known as the Floquet theorem, and it is a temporal version of Bloch's theorem for spatially periodic systems. We substitute Eq.~(\ref{eq:Flqthm}) into Eq.~(\ref{eq:Scheq}) and apply the Fourier transformations with respect to time. Then the time-dependent Schr\"{o}dinger equation is eventually rewritten as,
\end{comment}

Using the Floquet theory, problems of time-periodically driven systems can be effectively mapped onto problems of equilibrium states. Specifically, the time-dependent Schr\"odinger equation for the time-periodic Hamiltonian in Eq.~(\ref{eq:TBm2}) is rewritten in the form,
%%%%%%%%%%%%%%%%%%%%%%%%%
\begin{eqnarray}
\sum_{m=-\infty}^{\infty} \mathcal{H}_{nm} \ket{\Phi_\nu^m}
=\varepsilon^n_{\nu}\ket{\Phi_{\nu}^n},
\label{eq:H-Mw1}
\end{eqnarray}
%%%%%%%%%%%%%%%%%%%%%%%%%
where
%%%%%%%%%%%%%%%%%%%%%%%%%
\begin{eqnarray}
\mathcal{H}_{nm}=H_{n-m}-m\omega\delta_{n,m},
\label{eq:H-Mw2}
\end{eqnarray}
%%%%%%%%%%%%%%%%%%%%%%%%%
where $\mathcal{H}_{nm}$ denotes the matrix elements of Floquet matrix $\hat{\mathcal{H}}$. The Fourier coefficients $H_n$ and $\ket{\Phi_\nu^n}$ are defined by,
%%%%%%%%%%%%%%%%%%%%%%%%%
\begin{eqnarray}
\ket{\Phi_\nu^n}&=&\frac{1}{T}\int_0^T \ket{\Phi_{\nu}(\tau)}e^{in\omega \tau} d\tau,
\label{phiFC}\\
H_n&=&\frac{1}{T}\int_0^T H(\tau)e^{in\omega \tau} d\tau,
\label{HamFC}
\end{eqnarray}
%%%%%%%%%%%%%%%%%%%%%%%%% 
where the integers $n$ and $m$ correspond to the number of photons, and the index $\nu$ labels the eigenstates in each subspace of the photon number.

In the present study, we also use another approach based on an effective Hamiltonian, called Floquet effective Hamiltonian, in the high-frequency limit, which is derived by the $1/\omega$-expansion using the Brillouin-Wigner theorem~\cite{Mikami16}. The effective Hamiltonian is given by,
%%%%%%%%%%%%%%%%%%%%%%%%%
\begin{eqnarray}
H_{\mathrm{eff}}=H_0
+\sum_{n=1}^\infty \frac{[H_{-n}, H_n]}{n\hbar\omega}
+O\left(\frac{W^3}{\hbar^2\omega^2}\right),
\label{eq:Heff}
\end{eqnarray}
%%%%%%%%%%%%%%%%%%%%%%%%%
where $H_0$ is the 0th Fourier coefficient $H_{n=0}$. The Fourier coefficients of the time-periodic tight-binding Hamiltonian in Eq.~(\ref{eq:TBm2}) are calculated as,
%%%%%%%%%%%%%%%%%%%%%%%%%
\begin{eqnarray}
H_n &=& \sum_{\langle i,j \rangle}\sum_{\alpha, \beta} t_{i\alpha,j\beta} 
J_n(A_{i\alpha,j\beta})e^{-in\theta_{i\alpha,j\beta}}c_{i,\alpha}^{\dagger}c_{j,\beta}
\label{eq:HnLPL}
\end{eqnarray}
%%%%%%%%%%%%%%%%%%%%%%%%%
where $J_n$ is the $n$th Bessel function. Here, $A_{i\alpha,j\beta}$ and $\theta_{i\alpha,j\beta}$ are respectively defined as
%%%%%%%%%%%%%%%%%%%%%%%%%
\begin{align}
A_{i\alpha,j\beta}
%&=\frac{e}{\hbar}
%\left[\left\{bA_l(\tilde{x}_{i\alpha}-\tilde{x}_{j\beta})\cos\alpha
%     +aA_l(\tilde{y}_{i\alpha}-\tilde{y}_{j\beta})\sin\alpha\right\}^2\right. \nonumber \\
%     &+\left.\left\{-bA_s(\tilde{x}_{i\alpha}-\tilde{x}_{j\beta})\sin\alpha
%     +aA_s(\tilde{y}_{i\alpha}-\tilde{y}_{j\beta})\cos\alpha\right\}^2\right]^{1/2}
%\nonumber \\
&=
\left[\left\{\mathcal{A}_b(\tilde{x}_{i\alpha}-\tilde{x}_{j\beta})
     +\mathcal{A}_a(\tilde{y}_{i\alpha}-\tilde{y}_{j\beta})\right\}^2\right. \nonumber \\
     &+\left.\left\{-\mathcal{A}_d(\tilde{x}_{i\alpha}-\tilde{x}_{j\beta})
     +\mathcal{A}_c(\tilde{y}_{i\alpha}-\tilde{y}_{j\beta})\right\}^2\right]^{1/2}
     ,
\label{eq:Aij1} \\
\theta_{i\alpha,j\beta}
%&=\tan^{-1}\left[\frac{bA_l(\tilde{x}_{i\alpha}-\tilde{x}_{j\beta})\cos\alpha
%     +aA_l(\tilde{y}_{i\alpha}-\tilde{y}_{j\beta})\sin\alpha}{-bA_s(\tilde{x}_{i\alpha}-\tilde{x}_{j\beta})\sin\alpha
%     +aA_s(\tilde{y}_{i\alpha}-\tilde{y}_{j\beta})\cos\alpha}\right]
&=\tan^{-1}\left[\frac{\mathcal{A}_b(\tilde{x}_{i\alpha}-\tilde{x}_{j\beta})
     +\mathcal{A}_a(\tilde{y}_{i\alpha}-\tilde{y}_{j\beta})}{-\mathcal{A}_d(\tilde{x}_{i\alpha}-\tilde{x}_{j\beta})
     +\mathcal{A}_c(\tilde{y}_{i\alpha}-\tilde{y}_{j\beta})}\right],
\end{align}
%%%%%%%%%%%%%%%%%%%%%%%%%
with
%%%%%%%%%%%%%%%%%%%%%%%%%
\begin{eqnarray}
\mathcal{A}_a&=&\frac{eaA_{\rm \ell}\sin\alpha}{\hbar} \\
\mathcal{A}_b&=&\frac{ebA_{\rm \ell}\cos\alpha}{\hbar} \\
\mathcal{A}_c&=&\frac{eaA_{\rm s}\cos\alpha}{\hbar} \\
\mathcal{A}_d&=&\frac{ebA_{\rm s}\sin\alpha}{\hbar}.
\label{eq:Aij2}
\end{eqnarray}
%%%%%%%%%%%%%%%%%%%%%%%%%

After performing the Fourier transformations with respect to the spatial coordinates, we obtain,
%%%%%%%%%%%%%%%%%%%%%%%%%
\begin{eqnarray}
\hat{H}_n(\bm{k})=\left(
\begin{array}{cccc}
0 & A_{2,n}(\bm{k}) & B_{2,n}(\bm{k}) & B_{1,n}(\bm{k}) \\
A_{2,-n}^{*}(\bm{k}) & 0 & B_{2,-n}^{*}(\bm{k}) & B_{1,-n}^{*}(\bm{k}) \\
B_{2,-n}^{*}(\bm{k}) & B_{2,n}(\bm{k}) & 0 & A_{1,n}(\bm{k}) \\
B_{1,-n}^{*}(\bm{k}) & B_{1,n}(\bm{k}) & A_{1,n}(\bm{k}) & 0
\end{array}
\right),
\label{eq:Hk}
\end{eqnarray}
%%%%%%%%%%%%%%%%%%%%%%%%%
where
%%%%%%%%%%%%%%%%%%%%%%%%%
\begin{align}
A_{1,n}(\bm{k})&=
 t_{a1}\,\exp\left[i\frac{k_y}{2}\right]
J_{-n}(\sqrt{\mathcal{A}_a^2+\mathcal{A}_c^2}/2)e^{-in\theta} 
 \nonumber \\
&+t_{a1}\,\exp\left[-i\frac{k_y}{2}\right]
J_{n}(\sqrt{\mathcal{A}_a^2+\mathcal{A}_c^2}/2)e^{-in\theta},
\nonumber \\
A_{2,n}(\bm{k})&=
 t_{a2}\,\exp\left[i\frac{k_y}{2}\right]
J_{-n}(\sqrt{\mathcal{A}_a^2+\mathcal{A}_c^2}/2)e^{-in\theta}
 \nonumber \\
&+t_{a3}\,\exp\left[-i\frac{k_y}{2}\right]
J_{n}(\sqrt{\mathcal{A}_a^2+\mathcal{A}_c^2}/2)e^{-in\theta},
\nonumber \\
B_{1,n}(\bm{k})&=
 t_{b1}\,\exp\left[i\Bigl(\frac{k_x}{2}+\frac{k_y}{4}\Bigr)\right]
J_{-n}\left(\sqrt{\mathcal{A}_{\ell +}^2+\mathcal{A}_{{\rm s} -}^2}\right)e^{-in\theta_2}
 \nonumber \\
&\hspace{-0.3cm}
+t_{b4}\,\exp\left[-i\Bigl(\frac{k_x}{2}-\frac{k_y}{4}\Bigr)\right]
J_{-n}\left(\sqrt{\mathcal{A}_{\ell +}^2+\mathcal{A}_{{\rm s} -}^2}\right)e^{-in\theta_2},
\nonumber \\
B_{2,n}(\bm{k})&=
 t_{b2}\,\exp\left[i\Bigl(\frac{k_x}{2}-\frac{k_y}{4}\Bigr)\right]
J_{n}\left(\sqrt{\mathcal{A}_{\ell -}^2+\mathcal{A}_{{\rm s} +}^2}\right)e^{-in\theta_3}
\nonumber \\
&\hspace{-0.3cm}
+t_{b3}\,\exp\left[-i\Bigl(\frac{k_x}{2}+\frac{k_y}{4}\Bigr)\right]
J_{n}\left(\sqrt{\mathcal{A}_{\ell -}^2+\mathcal{A}_{{\rm s} +}^2}\right)e^{-in\theta_3},
\nonumber
\end{align}
%%%%%%%%%%%%%%%%%%%%%%%%%
with
%%%%%%%%%%%%%%%%%%%%%%%%%
\begin{align}
&\mathcal{A}_{\ell \pm}=\frac{eA_{\ell}}{4\hbar}(\pm2b\cos\alpha + a\sin\alpha), \\
&\mathcal{A}_{{\rm s} \pm}=\frac{eA_{\rm s}}{4\hbar}(\pm2b\cos\alpha + a\sin\alpha), \\
&\mathcal{\theta}=\tan^{-1}\left(\frac{A_{\ell}\sin\alpha}{A_{\rm s}\cos\alpha}\right), \\
&\mathcal{\theta}_{\pm}=\tan^{-1}\left[\frac{A_{\ell}(\pm2b\cos\alpha + a\sin\alpha)}{A_{\rm s}(\mp2b\cos\alpha + a\sin\alpha)}\right].
\label{eq:Aij3}
\end{align}
%%%%%%%%%%%%%%%%%%%%%%%%%
We use Eq.~(\ref{eq:Hk}) for $H_n$, $H_{-n}$ and $H_{n-m}$ in Eqs.~(\ref{eq:H-Mw2}) and (\ref{eq:Heff}). 

The Chern number of the $\nu$th band $N_{\rm Ch}^\nu$ ($\nu$=1-4) can be calculated from the Berry curvatures $B_z^{n\nu}(\bm k)$ as,
%%%%%%%%%%%%%%%%%%%%%%%%%%%%%%%%%%
\begin{eqnarray}
N_{\rm Ch}^\nu=\frac{1}{2\pi}\int\int_{\rm BZ}\;B_z^{0\nu}(\bm k) dk_xdk_y,
\end{eqnarray}
%%%%%%%%%%%%%%%%%%%%%%%%%%%%%%%%%%
where $B_z^{n\nu}(\bm k)$ is given by,
%%%%%%%%%%%%%%%%%%%%%%%%%%%%%%%%%%
\begin{align}
&B_z^{n\nu}(\bm k)=
\nonumber \\
&i\sum_{(m,\mu)}\frac{
\bra{\Phi_{\nu}^n(\bm k)}\frac{\partial \hat{\mathcal{H}}}{\partial k_x}\ket{\Phi_{\mu}^m(\bm k)}
\bra{\Phi_{\mu}^m(\bm k)}\frac{\partial \hat{\mathcal{H}}}{\partial k_y}\ket{\Phi_{\nu}^n(\bm k)}
-{c.c.}}
{[\varepsilon^m_\mu(\bm k)-\varepsilon^n_\nu(\bm k)]^2}.
\end{align}
%%%%%%%%%%%%%%%%%%%%%%%%%%%%%%%%%%
Here $\hat{\mathcal{H}}$ is the matrix of the Floquet Hamiltonian, while $\varepsilon^n_\nu(\bm k)$ and $\ket{\Phi_\nu^n(\bm k)}$ are the eigenenergies and eigenvectors of Eq.~(\ref{eq:H-Mw2}). The summation is taken over $m$ and $\mu$ where $(m,\mu)\ne(n,\nu)$; ``$c.c.$" is the complex conjugate of the first term of the numerator. In this work, we use a numerical method proposed by Fukui $et$ $al.$ in Ref.~\cite{Fukui05} to calculate the Chern numbers. Note that in the Chern insulator phase, the total Chern number $N_{\rm Ch}$ is defined as a sum of the Chern numbers $N_{\rm Ch}^\nu$ of bands below the Fermi level ($\nu$=1, 2, 3),
%%%%%%%%%%%%%%%%%%%%%%%%%
\begin{eqnarray}
N_{\rm Ch}=\sum_{\nu=1}^3 N_{\rm Ch}^\nu.
\end{eqnarray}
%%%%%%%%%%%%%%%%%%%%%%%%%
Because a sum over all the bands ($\nu$=1-4) is zero, the relation $N_{\rm Ch}=-N_{\rm Ch}^4$ holds.

Since the Floquet matrix is of infinite dimension, we consider a truncated Floquet matrix for practical calculations. More specifically, we restrict the number of photons to $|m|\leq 8$. Note that the ratio between the bandwidth $W$ and the light frequency $\hbar\omega$ determines the number of photons $m$ to be considered as $m\geq W/(\hbar\omega)$~\cite{Mikami16}. Because the typical bandwidth of $\alpha$-(BEDT-TTF)$_2$I$_3$ is $W\sim0.8$ eV, the restricted number of photons of $|m| \leq 8$ offers sufficiently accurate results when $\hbar \omega \gtrsim 0.1$ eV.

\section{Results}
%%%%%%%%%%%%%%%%%%%%%%%%%%%%%%%%
\begin{figure*}
\begin{center}
\includegraphics[scale=1.0]{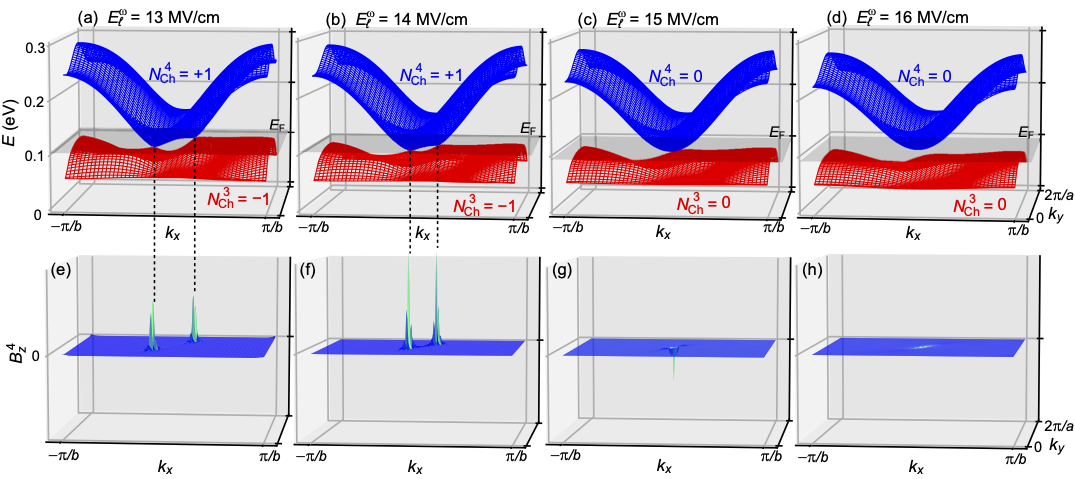}
\caption{(a)-(d) Quasienergy band structures in $\alpha$-(BEDT-TTF)$_2$I$_3$ irradiated by elliptically polarized light for different long-axis amplitudes $E_\ell^\omega$ of light, i.e., (a) $E_\ell^\omega=13$~MV/cm, (b) $E_\ell^\omega=14$~MV/cm, (c) $E_\ell^\omega=15$~MV/cm, and (d) $E_\ell^\omega=16$~MV/cm. (e)-(h) Berry curvatures for the fourth band in respective band structures. When a weak light field is applied, a pair of gapped Dirac points with positive Berry-curvature peaks appear, indicating the emergence of photoinduced topological phase. The distance between the two Dirac points becomes closer as $E_\ell^\omega$ increases, and they eventually collide to annihilate when $E_\ell^\omega=15$~MV/cm. At $E_\ell^\omega=16$~MV/cm, a gapped band structure with zero Berry curvature appears, indicating the occurence of photinduced phase tranition to a nontopological insulator phase. The frequency, polarization angle, and short-axis amplitude of elliptically polarized light are fixed at $\hbar\omega=0.6$~eV, $\alpha=45^\circ$, and $E^\omega_{\rm s}=2$~MV/cm, respectively.}
\label{Fig03}
\end{center}
\end{figure*}
%%%%%%%%%%%%%%%%%%%%%%%%%%%%%%%%

We first discuss the photoinduced collision and collapse of two massive Dirac points and resulting novel topological phase transition in $\alpha$-(BEDT-TTF)$_2$I$_3$. Figures~\ref{Fig03}(a)-(d) show quasienergy band structures under elliptically polarized light for different long-axis amplitudes $E^\omega_{\rm \ell}$ of light, i.e., (a) $E^\omega_{\rm \ell}$=13 MV/cm, (b) $E^\omega_{\rm \ell}$=14 MV/cm, (c) $E^\omega_{\rm \ell}$=15 MV/cm, and (d) $E^\omega_{\rm \ell}$=16 MV/cm, where the other parameters of the light are chosen as $\hbar\omega$=0.6 eV, $\alpha=45^\circ$, and $E^\omega_{\rm s}$=2 MV/cm. These results are obtained by diagonalizing the Floquet Hamiltonian in Eq.~(\ref{eq:H-Mw2}). In Figs.~\ref{Fig03}(e)-(h), the Berry curvatures of the fourth band for corresponding quasienergy band structures are presented. 

Since elliptically polarized light breaks the time-reversal symmetry, it opens a topological gap at the two Dirac points that are initially gapless, and brings about a phase transition to a topologically nontrivial phase. As shown in Figs.~\ref{Fig03}(a) and (b), the quasienergy bands of this topological phase cross the Fermi level at certain momentum points distinct from those of the gapped Dirac points, indicating that the system attains a metallic conductivity. Therefore, we call this phase as the topological semimetal with a pair of gapped Dirac cones. The Chern numbers of the third and fourth bands are $N_{\rm Ch}^3=-1$ and $N_{\rm Ch}^4=+1$, respectively. As shown in Figs.~\ref{Fig03}(e) and (f), the Berry curvature of the fourth band has positive peaks at the two Dirac points. These positive Berry curvatures give rise to nonzero Hall conductivity, which will be discussed later. This photoinduced topological phase transition is nothing but the one that has been intensively studied since its prediction for a photodriven Dirac-electron system in graphene~\cite{Oka09, Inoue10, Mikami16, Kitayama20, Kitayama21b}.

As we increase the long-axis amplitude $E^\omega_{\rm \ell}$ of light, these two Dirac points get closer in the momentum space [Figs.~\ref{Fig03}(a) and (b), and Figs.~\ref{Fig03}(e) and (f)]. This approaching behavior of the Dirac points is caused by the band deformation due to the photoinduced renormalization of transfer integrals as discussed in Ref.~\cite{Kitayama21a}. When $E^\omega_{\rm \ell}$ reaches $\sim$15 MV/cm, these Dirac points collide [Figs.~\ref{Fig03}(c) and (g)], and eventually disappear [Figs.~\ref{Fig03}(d) and (h)]. After the disappearance of Dirac points, the quasienergy band structure still has a gap that separates the third and fourth bands as seen in Fig.~\ref{Fig03}(d). However, the system is no longer topological, that is, the Chern numbers of the third and fourth bands are both zero, indicating the emergence of nontopological phase under irradiation with relatively intense elliptically polarized light of $E^\omega_{\rm \ell} \gtrsim 15$ MV/cm. This phenomenon was studied before for the linearly polarized light~\cite{Kitayama21a}. The present result is an extension to the case of elliptically polarized light with $E^\omega_{\rm \ell} \gg E^\omega_{\rm s}$. Note that this photoinduced phase transition is a transition from topological to nontopological phases and originates from a novel physical mechanism distinct from the usually argued mechanism based on the time-reversal symmetry breaking.

%%%%%%%%%%%%%%%%%%%%%%%%%%%%%%%%
\begin{figure}
\begin{center}
\includegraphics[scale=1.0]{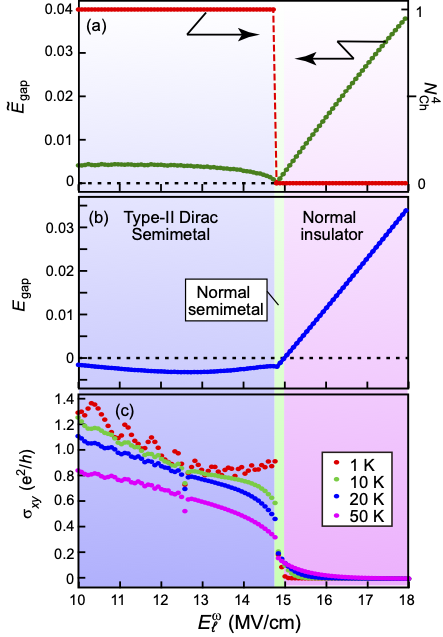}
\caption{Calculated dependencies of several physical quantities, i.e., (a) the band gap $\tilde{E}_{\rm gap}$, the Chern number of the fourth band $N_{\rm ch}^4$, (b) the energy gap $E_{\rm gap}$, and (c) the Hall conductivities $\sigma_{xy}$ at various temperatures, on the long-axis amplitude $E^\omega_{\rm \ell}$ of light, which characterize the predicted novel photoinduced topological-to-nontopological phase transition with collision and collapse of the Dirac points in $\alpha$-(BEDT-TTF)$_2$I$_3$ irradiated with elliptically polarized light. The light parameters are fixed at $\hbar\omega$=0.6 eV, $\alpha=45^\circ$, and $E_{\rm s}^\omega$=2 MV/cm, respectively.}
\label{Fig04}
\end{center}
\end{figure}
%%%%%%%%%%%%%%%%%%%%%%%%%%%%%%%%
To see unusual aspects of this novel photoinduced topological phase transition, we calculate dependencies of several physical quantities on the long-axis amplitude $E^\omega_{\rm \ell}$ of light [Fig.~\ref{Fig04}]. The profile of the Chern number $N_{\rm Ch}^4$ in Fig.~\ref{Fig04}(a) shows an abrupt change from $+1$ to 0 at $E^\omega_{\rm \ell}$=14.75 MV/cm, indicating the occurence of photoinduced phase transition from topological to nontopological phases. Upon this change in the band topologies, the gap between the third and fourth bands is required to close once. This can be seen in the profile of band gap $\tilde{E}_{\rm gap}$, which is defined by,
%%%%%%%%%%%%%%%%%%%%%%%%%
\begin{eqnarray}
\tilde{E}_{\rm gap} = \min[\varepsilon^0_4(\bm{k}) - \varepsilon^0_3(\bm{k})].
\label{eq:Egap2}
\end{eqnarray}
%%%%%%%%%%%%%%%%%%%%%%%%%
This quantity is the minimum energy distance between the third and fourth bands. If the Dirac cones exist, it corresponds to the magnitude of gap at the Dirac points. Therefore, this quantity can be used to judge whether the third and fourth bands touch at some points in the momentum space. More specifically, $\tilde{E}_{\rm gap}>0$ indicates that the two bands are completely separated over the entire momentum space even at the Dirac points, if any, whereas $\tilde{E}_{\rm gap}=0$ indicates that they touch at some momentum points. The profile of $\tilde{E}_{\rm gap}$ shown in Fig.~\ref{Fig04}(a) clearly indicates the closing of band gap at the point where the Chern number changes from $N_{\rm Ch}^4$=+1 to $N_{\rm Ch}^4$=0.

We also introduce another quantity called energy gap $E_{\rm gap}$ defined by,
%%%%%%%%%%%%%%%%%%%%%%%%%
\begin{eqnarray}
E_{\rm gap} = \min[\varepsilon^0_4(\bm{k})] - \max[\varepsilon^0_3(\bm{k})].
\label{eq:Egap1}
\end{eqnarray}
%%%%%%%%%%%%%%%%%%%%%%%%%
This quantity can be exploited to judge whether the system is semimetal or insulator. When $E_{\rm gap}<0$, the top of third band is located at higher energy than the bottom of fourth band. In this situation, the Fermi level necessarily runs across these bands, and the system attains metallic conductivity. On the other hand, when $E_{\rm gap}>0$, the top of the third band is located below the fourth band. In this situation, the Fermi level runs within a band gap between these two bands without running across the bands, and the system is insulating. Figure~\ref{Fig04}(b) shows the calculated profile of $E_{\rm gap}$. According to this figure as well as Fig.~\ref{Fig04}(a), we find that a normal semimetal phase ($\tilde{E}_{\rm gap}>0$ and $E_{\rm gap}<0$) emerges in a tiny region of 14.75 MV/cm $<E^\omega_{\rm \ell}<$15 MV/cm next to the Dirac semimetal phase. This phase is nontopological with a vanishing Chern number $N_{\rm Ch}^4$=0 and has a metallic conductivity with $E_{\rm gap}<0$. In the subsequent region of $E^\omega_{\rm \ell}>$15 MV/cm, the normal insulator phase appears with $N_{\rm Ch}^4$=0 and $E_{\rm gap}>0$.

The present photoinduced topological phase transition can be seen in the profiles of Hall conductivity $\sigma_{xy}$ as well. The Hall conductivity in the photoinduced nonequilibrium phases can be calculated by using the formula,
%%%%%%%%%%%%%%%%%%%%%%%%%
\begin{eqnarray}
\sigma_{xy} = \frac{2e^2}{h} \int_{\rm BZ} \frac{dk_x dk_y}{2\pi} \sum_{\nu=1}^4 n_{\nu}(\bm{k}) B_z^\nu(\bm{k}).
\label{eq:FlqTKNN}
\end{eqnarray}
%%%%%%%%%%%%%%%%%%%%%%%%%
Here $n_{\nu}(\bm{k})$ is the nonequilibrium distribution function of the $\nu$th Floquet band in the zero-photon subspace, which is calculated using the Floquet-Keldysh formalism~\cite{Oka09,Kitayama20,Kitayama21b}. The factor 2 comes from the spin degeneracy. Figure~\ref{Fig04}(c) shows the calculated $E^\omega_{\rm \ell}$ dependencies of $\sigma_{xy}$ at various temperatures. This quantity takes large values in the topological semimetal phase with $N_{\rm Ch}^4$=1, but it decreases abruptly with a jump at the transition point to the normal semimetal phase with $N_{\rm Ch}^4$=0. In the normal insulator phase with $N_{\rm Ch}^4$=0, this quantity is almost suppressed to be zero. These results indicate that the predicted successive photoinduced phase transitions might be experimentally detected by measurement of the Hall conductivity under the photoirradiation.

%%%%%%%%%%%%%%%%%%%%%%%%%%%%%%%%
\begin{figure}[tb]
\begin{center}
\includegraphics[scale=1.0]{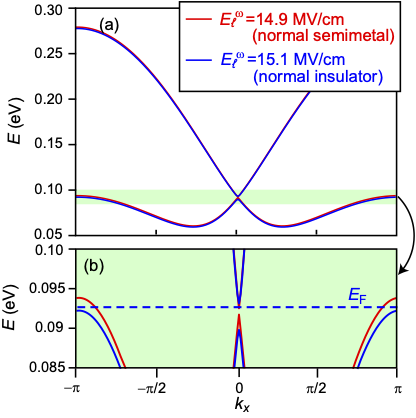}
\caption{(a) Qusienergy band structures along $k_x$ at $k_y=\pi/a$ of $\alpha$-(BEDT-TTF)$_2$I$_3$ irradiated with elliptically polarized light for different long-axis amplitudes of light, i.e., $E^\omega_{\rm \ell}$=14.9 MV/cm (normal semimetal phase) and  $E^\omega_{\rm \ell}$=15.1 MV/cm (normal insulator phase). (b) Magnified view of the band structure near the Fermi level. The light parameters are fixed at $\hbar\omega$=0.6 eV, $\alpha=45^\circ$, and $E^\omega_{\rm s}$=2 MV/cm.}
\label{Fig05}
\end{center}
\end{figure}
%%%%%%%%%%%%%%%%%%%%%%%%%%%%%%%%
The emergence of normal semimetal phase in a tiny region between the topological semimetal phase and the normal insulator phase, i.e., 14.75 MV/cm $<E^\omega_{\rm \ell}<$15 MV/cm, can be discussed by focusing on the fine band structures. Figure~\ref{Fig05}(a) presents the quasienergy band structures along ($k_x$, $\pi/a$) for $E^\omega_{\rm \ell}$=14.9 MV/cm (normal semimetal phase) and $E^\omega_{\rm \ell}$=15.1 MV/cm (normal insulator phase), while Fig.~\ref{Fig05}(b) magnifies those near the Fermi level. Both phases have finite band gaps $\tilde{E}_{\rm gap}>0$ [Fig.~\ref{Fig04}(a)], manifested by a gap at $k_x=0$. On the contrary, the sign of the energy gap $E_{\rm gap}$ is opposite between these two phases. The normal semimetal phase has a negative energy gap $E_{\rm gap}<0$. This means that the Fermi level runs over the third and fourth bands, and thus the system becomes a semimetal with metallic conductivity. On the other hand, the normal insulator phase has a positive energy gap $E_{\rm gap}>0$. This means that the Fermi level runs within a gap between the third and fourth bands and does not cross these bands. Consequently, the system attains insulating nature. As seen in Fig.~\ref{Fig05}(b), the third band crosses the Fermi level at $k_x=\pm \pi/b$ for the normal semimetal phase at $E^\omega_{\rm \ell}$=14.9 MV/cm, whereas it does not for the normal insulator phase at $E^\omega_{\rm \ell}$=15.1 MV/cm. In Fig.~\ref{Fig02}, we have summarized four possible band structures in the present system.

%%%%%%%%%%%%%%%%%%%%%%%%%%%%%%%%
\begin{figure}[tb]
\begin{center}
\includegraphics[scale=1.0]{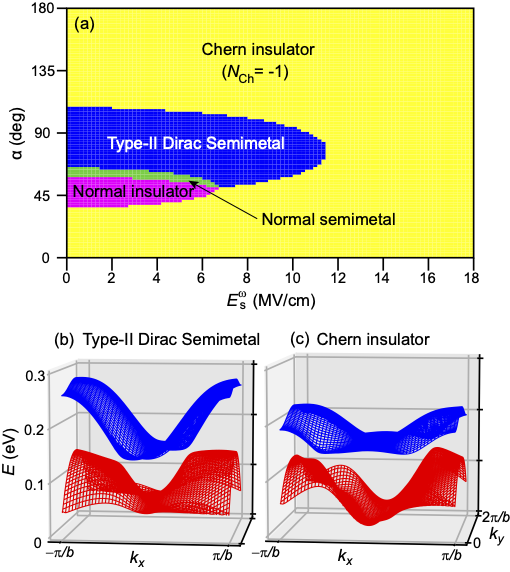}
\caption{(a) Nonequilibrium phase diagram of $\alpha$-(BEDT-TTF)$_2$I$_3$ under irradiation with elliptically polarized light in the plane of the short-axis amplitude $E^\omega_{\rm s}$ and the polarization angle $\alpha$ of light when $\hbar\omega$=0.67 eV and $E^\omega_{\rm \ell}$=18 MV/cm. The topological semimetal phase has gapped Dirac points. (b), (c) Typical quasienergy band structures of (b) the topological semimetal phase and (c) the Chern insulator phase, which are calculated for $\alpha=90^\circ$ and $\alpha=135^\circ$, respectively. The light parameters are fixed at $\hbar\omega$=0.67 eV, $E^\omega_{\rm s}$=7 MV/cm, and $E^\omega_{\rm \ell}$=18 MV/cm for both cases.}
\label{Fig06}
\end{center}
\end{figure}
%%%%%%%%%%%%%%%%%%%%%%%%%%%%%%%%
Finally, we construct a nonequilibrium phase diagram of $\alpha$-(BEDT-TTF)$_2$I$_3$ driven by elliptically polarized light by the physical quantities $\tilde{E}_{\rm gap}$, $E_{\rm gap}$, and $N_{\rm ch}^4$. Figure~\ref{Fig06}(a) shows the phase diagram in plane of the short-axis amplitude $E_{\rm s}^\omega$ and the polarization angle $\alpha$. Here the long-axis amplitude and the frequency of light are fixed at $E^\omega_{\rm \ell}$=18 MV/cm and $\hbar\omega$=0.67 eV, respectively. In the limit of $E_{\rm s}^\omega$=0, the light is linearly polarized. In this case, our previous work in Ref.~\cite{Kitayama21a} predicted that a pair annihilation of Dirac points occurs owing to the deformation of band structure due to the dynamical renormalization of transfer integrals when $\alpha\sim 45^\circ$. This Dirac-point annihilation leads to the emergence of normal insulator phase as seen in the phase diagram at $\alpha \sim 45^\circ$ and $E_{\rm s}^\omega$=0. This normal insulator phase remains even if the light is not perfectly of linear polarization but elliptically polarized with nonzero $E_{\rm s}^\omega$ as long as the polarization angle $\alpha$ is nearly $45^\circ$. The phase diagram indicates that the normal insulator phase survives up to $E^\omega_{\rm \ell} \sim 7$ MV/cm, which corresponds to the ellipticity of $E_{\rm s}^\omega/E^\omega_{\rm \ell} \sim 0.39$. On the other hand, when $E_{\rm s}^\omega \gtrsim$12 MV/cm, the system is lying in the Chern insulator phase irrespective of the polarization angle $\alpha$. Note that when $E_{\rm s}^\omega$=18 MV/cm, the light is of perfect circular polarization with $E_{\rm s}^\omega=E^\omega_{\rm \ell}$.

When the organic salt $\alpha$-(BEDT-TTF)$_2$I$_3$ is irradiated by elliptically polarized light with nonzero $E^\omega_{\rm s}$, the Dirac points, if any, must be gapped because of the photoinduced breaking of time-reversal symmetry. Thereby, as long as the Dirac points exist, either the topological semimetal phase or the Chern insulator phase emerges. The topological semimetal phase has the bands crossing the Fermi level although the Dirac points are gapped, which carry the metallic conductivity. On the contrary, the Fermi level is located bewteen the well-separated third and fourth bands without crossing them in the Chern insulator phase, which renders the bulk insulating. Figures~\ref{Fig06}(b) and (c) show typical quaienergy band structures for the topological semimetal phase and the Chern insulator phase under irradiation by elliptically polarized light. Note that the quasienergy band structure in the Chern insulator phase in Fig.~\ref{Fig06}(c) is significantly deformed from the original band structue at equilibrium [Fig.~\ref{Fig01}(b)]. This exemplifies the photoinduced band deformation due to the dynamical renormalization of transfer integrals. 

\section{Summary}
In summary, we have theoretically predicted possible novel type of photoinduced topological phase transition in $\alpha$-(BEDT-TTF)$_2$I$_3$ that possesses a pair of Dirac-cone bands at equilibrium. By constructing a Floquet theory for this organic compound driven by elliptically polarized light, we demonstrate that the irradiation with a specified elliptical-axis angle of light causes the collision of two massive Dirac points and their collapse through the photoinduced band deformation, which eventually results in the phase transition from the topological semimetal phase to the nontopological phases. The photoinduced topological phase transitions have been intensively studied since its theoretical prediction in photodriven graphene, but most of the previous studies have dealt with phenomena based basically on the same physical mechanism, i.e., gap opening at the Dirac-electron bands by the light-induced time-reversal symmetry breaking. On the contrary, the phenomenon predicted in this work is based on a totally different mechanism, and thus is novel. We have also discussed that the Hall conductivity can be a good probe for experimental detection of this phase transition. Our work has added a new fundamental physics to the optical control of topologies in matters and thus has provided significant advances to this important research field.

\section{Acknowledgment}
This work was supported by JSPS KAKENHI (Grant No. 17H02924, No. 16H06345, No. 18H01162, No. 19K23427, No. 20K03841, and No. 21J20856) and Waseda University Grant for Special Research Projects (Project No. 2021C-566).

\end{document}